# Cooperative Cognitive Radio Network with Energy Harvesting: Stability Analysis

Ramy Amer [1], Amr A. El-Sherif [2], Hanaa Ebrahim [1] and Amr Mokhtar [2]

[1] Switching Department, National Telecommunication Institute, Cairo, Egypt.

[2] Dept. of Electrical Engineering, Alexandria University, Alexandria 21544, Egypt.

{rami.amer@nti.sci.eg, aasherif@alexu.edu.eg, dr.hanaa_nti@yahoo.com, amromokhtar61@gmail.com}

*Abstract*— This paper investigates the maximum stable throughput of a cooperative cognitive radio system with energy harvesting Primary User (PU) and Secondary User (SU). Each PU and SU has a data queue for data storage and a battery for energy storage. These batteries harvest energy from the environment and store it for data transmission in next time slots. The SU is allowed to access the PU channel only when the PU is idle. The SU cooperates with the PU for its data transmission, getting mutual benefits for both users, such that, the PU exploits the SU power to relay a fraction of its undelivered packets, and the SU gets more opportunities to access idle time slots. To characterize the system's stable throughput region, it is noted that the queues in the system are interacting, i.e., the service process of any queue depends on the current state of the other queues, which renders the analysis intractable. To simplify the analysis, a dominant system approach is used to obtain a closed form expressions for the system's stable throughput region. Results reveal that, the non-cooperative system outperforms the cooperative system for low SU energy harvesting rate and irrespective of the PU energy harvesting rate, while the cooperation benefits are seen for high SU energy harvesting rate.

*Keywords*— relay; dominant system; energy harvesting; stable throughput region

## I. INTRODUCTION

Secondary utilization of a licensed spectrum band can enhance the spectrum usage and introduce a reliable solution to its scarcity. Secondary users (SUs) can access the spectrum under the constraint that a minimum quality of service is guaranteed for the primary users (PUs) [1]. In order to achieve cognitive radio objectives, SUs are required to adaptively modify its transmission parameters and to access radio spectrum without causing severe interference to the PU.

Recently, cooperation between the SU and PU has gained a lot of attention in cognitive radio research. Specifically, SUs act as relays for the PU data while also trying to transmit their own data. In [2], the advantages of the cognitive transmitter acting as a "transparent relay" for the PU transmission are investigated. The authors proved that the stability region of the system increases in terms of the maximum allowed arrival rates of both the PU and SU. Moreover, it was shown that the maximum allowed transmission power for the SU increases. The stability of PU and SU queues and throughput of a two-user cognitive radio system with multicast traffic is discussed in [3], where one node could acts as a relay for the packets of the other node's failed packets. It is shown that the stable throughput region of this cooperative system is larger than that of its non-cooperative counterpart. In [4] the protocol design for cognitive cooperative systems with many secondary users is proposed. In contrast with previous cognitive configurations, the channel model considered assumes a cluster of secondary users which perform both a sensing process for transmitting opportunities and can relay data for the primary user.

Energy limitations and constraints on transmission power have recently gained a lot of interest, specifically in cognitive radio systems. Energy harvesting has appeared as an alternate power supply, where each node harvests energy from the surrounding environment. Several articles have discussed energy harvesting solution for hard-wiring or replacing the batteries of rechargeable wireless devices [5], [6], [7].

Non-cooperative energy harvesting cognitive radio network with the general multipacket reception channel model, where the primary transmission may succeed even in the presence of secondary transmission, is investigated in [8]. Thus the cognitive user can increase its throughput through not only utilizing the idle periods of the primary user also randomly accessing the channel by some probability. First, the SU is assumed to harvest energy for transmission, and then, both the PU and SU are assumed to be equipped with rechargeable batteries. In [9], the effects of network-layer cooperation in a wireless three-node network with energy harvesting nodes are studied. Energy harvesting is modeled in each node as a buffer that stores the harvested energy. In [10], a system consisting of one PU and one SU, where the SU is harvesting energy from the ambient radio environment and follows a save-then-transmit method is investigated. Authors in [11] studied the queues stability in a slotted ALOHA random access network in which two nodes have finite energy sources. The two nodes have a battery for energy storage. Each node is modeled with two queues, the first for storing packets and the second models the energy in the battery. The stability region obtained is compared with the stability region of the system without energy constraints.

In this paper, we study the randomized cooperative policy with energy harvesting PU and SU nodes. The stable throughput region of the system is characterized for different PU and SU energy harvesting rates. Moreover, the energy constrained cooperative system is compared with the cooperative system without energy constraints, and the non-cooperative energy constrained system. Furthermore, we characterize the conditions for the system to switch between cooperative and non-cooperative modes. To the best of our knowledge, the problem of characterizing the stable throughput region of the cooperative PU and SU with energy harvesting at both nodes has not been studied before.

The rest of this paper is organized as follows. Section II introduces the system model. Stability conditions are found in Section III. Section IV presents the numerical results comparing the stability regions of the different systems under investigation. Finally, concluding remarks are drawn in Section V.

## II. SYSTEM MODEL

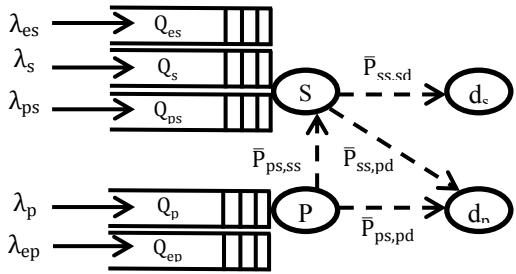

Fig. 1. System model.

Fig. 1 depicts the system model under consideration. The system is composed of one PU and one SU, assumed to harvest energy from the environment. PU has two queues, $Q_p$ and $Q_{ep}$. $Q_p$ is an infinite capacity buffer for storing the PU's fixed length packets. The arrival process at $Q_p$ is modeled as Bernoulli arrival process with mean $\lambda_p$ [packets/slot]. $Q_{ep}$ models the PU's battery, assumed to have an infinite size to store the harvested energy. Energy is assumed to be harvested in a certain unit and one unit of energy is consumed in each transmission attempt, assuming one unit of energy is equal to the transmission power of the source multiplied by the time of packet transmission. The energy harvesting process is modeled as a Bernoulli arrival process with mean $\lambda_{ep}$. These processes are independent, stationary and identically distributed (i.i.d) over time slots. Considering the SU, it is represented by three queues: $Q_s$, $Q_{ps}$, and $Q_{es}$. $Q_s$ is an infinite capacity buffer for storing the SU's own packets. The secondary relay queue, $Q_{ps}$, stores the PU's packets successfully received by the SU when the channel between the PU transmitter and receiver is in outage. $Q_{es}$ is the SU battery of infinite size storing the harvested energy. The arrival processes at the two queues, $Q_s$ and $Q_{es}$, are modeled as Bernoulli arrival process with means $\lambda_s$ and $\lambda_{es}$, respectively. Time is slotted, and a packet transmission takes one time slot. Therefore, the average arrival rates $\lambda_p$ and $\lambda_s$ [packets/slot] lie in the interval [0, 1]. The arrival processes at each user are independent and identically distributed across successive time slots (i.i.d). The average arrival rates $\lambda_{ep}$ and $\lambda_{es}$ [energy packets/slot] lie in the interval [0, 1].

The PU transmits a packet from $Q_p$ whenever it is non-empty. If the channel between the PU transmitter and receiver is not in outage, then the PU receiver successfully decodes the packet and the packet departs the system. It is assumed that the SU can overhear the ACK/NACK from the PU receiver. In the time slots where the channel between the PU transmitter and receiver is in outage, if the SU received the PU packet correctly, the packet will be stored in the relay queue and the SU will bear the responsibility to deliver this packet. If the channel between the PU and SU also is in outage, PU will try to retransmit the packet in a subsequent time slot. Any transmission or retransmission from $Q_p$ requires that $Q_{ep}$ be non-empty. The SU is assumed to perform perfect sensing. Whenever the channel is sensed to be idle, the secondary has two data queues to transmit a packet, specifically $Q_s$ and $Q_{ps}$. The SU is assumed to transmit a packet from $Q_s$ with probability $a$, or from $Q_{ps}$ with the complement probability $\bar{a} = 1 - a$.

## III. ENERGY HARVESTING STABLE THROUGHPUT REGIONS

In this section, the stable throughput region of the system under consideration is characterized. This region is bounded by the maximum arrival rates at the PU and SU when the two queues, $Q_p$, $Q_s$ are stable. The stability of the queue is identified by Loyne's theorem [12]. The theorem states that if the arrival and service process are stationary, then the queue is stable if the condition that the arrival rate is strictly less than the service rate is satisfied. For any queue in the system, the stability requires that:

$$\lambda_i < \mu_i, \tag{1}$$

where i = {p, ps, s, ep, es}, and $\mu_i$ refers to the service rate of the i$^{\text{th}}$ queue.

Starting with the PU data queue stability, a packet is serviced from $Q_p$ if it is successfully decoded by the PU destination, or by the SU. Let $\bar{P}_{j,k}$ denote the probability that the channel is not in outage between j and k, where j = {ps, ss}, k = {pd, ss, sd}, also ps, ss, pd, and sd represent the PU source, SU source, PU destination, and SU destination, respectively. We then have,

$$\mu_p = (\bar{P}_{ps,pd} + P_{ps,pd}\bar{P}_{ps,ss}) \Pr\{Q_{ep} \neq 0\}. \tag{2}$$

The probability that $Q_{ep}$ is empty is obtained from the Little's law, [13], by $(1 - \lambda_{ep}/\mu_{ep})$, where $\lambda_{ep}$ and $\mu_{ep}$ denotes the arrival and service rate of $Q_{ep}$, respectively. It is obvious that the service rate of the PU battery queue $Q_{ep}$ depends on whether the PU data queue $Q_p$ is empty or not. Similarly, the service rate of $Q_p$ depends on the state of $Q_{ep}$. This interdependence between the two queues results in an interacting system of queues. To decouple this interaction and simplify the analysis, we assume that $Q_p$ is saturated to formulate an expression for the service rate of $Q_{ep}$. The PU is assumed to always have a packet to transmit; this implies that each time slot an energy packet is consumed from $Q_{ep}$. So, the $Q_{ep}$ service rate, $\mu_{ep} = 1$. So, the probability that $Q_{ep}$ is not empty is $\lambda_{ep}/1$, and the probability that $Q_{ep}$ is empty is $(1 - \lambda_{ep})$. Substituting in (2), gives:

$$\mu_p = (\overline{P}_{ps,pd} + P_{ps,pd}\overline{P}_{ps,ss}) \lambda_{ep}. \quad (3)$$

The resulting PU's service rate under the saturation assumption is a lower bound on the actual service rate, therefore the obtained stability region will be an inner bound to the actual stability region.

For the relay queue at the SU, $Q_{ps}$, a packet from the PU enters the relay queue when the channel between the PU transmitter and receiver is in outage, the channel is not in outage between the PU and the SU, the PU battery is not empty, and the PU data queue is not empty, therefore,

$$\lambda_{ps} = P_{ps,pd}\overline{P}_{ps,ss} \lambda_{ep} \frac{\lambda_p}{\mu_p}. \quad (4)$$

The probability that the PU is idle is denoted by $I$. the PU is active when both the data queue and the battery queues are non-empty together, otherwise, the PU is idle, hence,

$$I = 1 - \lambda_{ep} (\lambda_p/\mu_p). \quad (5)$$

In our model, randomized cooperative policy, the SU transmits a packet from $Q_s$ or $Q_{ps}$ with probabilities $a$ and $\bar{a}$, respectively. In [14], a comparison between the literature cooperative model, in which a full priority is given to the relay queue, and the randomized cooperative policy is illustrated. It was shown that, the randomized cooperative policy enhanced the SU delay at the expense of a slight degradation in the PU delay. So, we chose the randomized cooperative policy as the cooperation model between the two energy harvesting primary and secondary users. A packet is serviced from $Q_{ps}$, with a probability $\bar{a}$ if the SU data queue, $Q_s$, is non-empty, or with a probability 1 if the SU data queue, $Q_s$, is empty (work conserving system). $\mu_{ps}$ can then be expressed as,

$$\mu_{ps} = \overline{P}_{ss,pd} \Pr(Q_{es} \neq 0) I \{ \bar{a}\Pr(Q_s \neq 0) + (1)\Pr(Q_s = 0)\}, \quad (6)$$

where $\Pr(Q_s = 0) = 1 - \lambda_s/\mu_s$.

Similarly, a packet is serviced from $Q_s$, with a probability $a$ if the SU relay queue, $Q_{ps}$, is non-empty, or with a probability 1 if the SU relay queue, $Q_{ps}$, is empty. $\mu_s$ can then be expressed as,

$$\mu_s = \overline{P}_{ss,sd} \Pr(Q_{es} \neq 0) I \{ a\Pr(Q_{ps} \neq 0) + (1)\Pr(Q_{ps} = 0)\}. \quad (7)$$

From (6) and (7), the service rate of the relay queue, $Q_{ps}$, depends on the current state of the SU data queue, $Q_s$, and the service rate of the SU data queue, $Q_s$, depends on the current state of the relay queue, $Q_{ps}$. Therefore, the two queues are interacting and the individual departure processes cannot be computed directly. So, we resort to the dominant system approach [15] [16] to decouple this interaction. In [17], characterization of the stability region of the slotted ALOHA for the two-node case over a collision channel when nodes are subject to energy availability constraints imposed by the battery status depended on the stochastic dominance technique.

A dominant system has the property that it is stable if and only if the original system is stable, and that its queues are not interacting. The dominant system can be determined by this simple modification to the original system: if $Q_{ps}$ (or $Q_s$) is empty; the SU continues to transmit "dummy" packets whenever it senses the PU is idle. If the SU transmits a dummy packet from $Q_s$ (dominant system *I*), then, $\Pr(Q_s \neq 0) = 1$ and $\Pr(Q_s = 0) = 0$, so from (6), a packet is transmitted from $Q_{ps}$ with probability $\bar{a}$ regardless of the actual state of $Q_s$. Conversely, if the SU transmits a dummy packet from $Q_{ps}$ (dominant system *II*), then, $\Pr(Q_{ps} \neq 0) = 1$ and $\Pr(Q_{ps} = 0) = 0$, so from (7), a packet is transmitted from $Q_s$ with probability $a$ regardless the actual state of $Q_{ps}$. So, in the two dominant systems, $Q_s$ and $Q_{ps}$ are decoupled and the service rates of $Q_s$ and $Q_{ps}$ could be computed directly. The stable throughput region of the original system would be the union of stable throughput region of the two dominant systems

For a packet to be transmitted from the SU data queues ($Q_{ps}$ or $Q_s$), it is served by an energy packet from $Q_{es}$. The probability of $Q_{es}$ being non-empty is $\lambda_{es}/\mu_{es}$. Since the SU is assumed to transmit dummy packets (from $Q_{ps}$ or $Q_s$), the service rate, $\mu_{es}$, of $Q_{es}$ is the probability that the PU is idle. The probability of $Q_{es}$ being non-empty is $\lambda_{es}/\mu_{es}$, and

$$\Pr(Q_{es} \neq 0) = \lambda_{es}/(1 - \lambda_{ep}(\lambda_p/\mu_p)). \quad (8)$$

Here, the individual departure processes will be computed in the two dominant systems. The stability of the queues under the two dominant systems will be investigated in the next two sections, and the two stability regions of the two systems will be expressed, stability region (*I*) and region (*II*). The stability of the original system is the union of stability region (*I*) and region (*II*).

*A. Dominant System (I)*

The SU is assumed to transmit dummy packets from $Q_s$, so the service rate of the relay queue, $Q_{ps}$, is independent of the state of $Q_s$. A packet is served from $Q_{ps}$ with probability $\bar{a}$ if the PU is idle, the channel between SU source and the PU destination is not in outage, and $Q_{es}$ not empty, therefore,

$$\mu_{ps} = \overline{P}_{ss,pd} \frac{\lambda_{es}}{1 - \lambda_{ep} (\lambda_p / \mu_p)} I \bar{a}. \quad (9)$$

For the SU relay queue stability, equation (1) requires that

$$P_{ps,pd}\overline{P}_{ps,ss} \lambda_{ep} \frac{\lambda_p}{\mu_p} < \overline{P}_{ss,pd} \frac{\lambda_{es}}{1 - \lambda_{ep} (\lambda_p / \mu_p)} I \bar{a},$$

$$\lambda_p < \frac{\bar{a} \, \overline{P}_{ss,pd}}{P_{ps,pd}\overline{P}_{ps,ss} \lambda_{ep}} \frac{\lambda_{es}}{1 - \lambda_{ep} (\lambda_p / \mu_p)} I \, \mu_p. \quad (10)$$

It is clear from the above relations that the maximum allowable arrival rate at $Q_p$ increases as $\bar{a}$ increases, which means that the SU increases the probability of serving the relay queue $Q_{ps}$. Another important note is that this increase implies that the SU has enough energy packets to serve the relay queue. This interprets (10) which show that $\lambda_p$ increases as $\lambda_{es}$ increases. From (7) and with the relations: $\Pr(Q_{ps} \neq 0) = \lambda_{ps}/\mu_{ps}$ and $\Pr(Q_{ps} = 0) = 1 - \lambda_{ps}/\mu_{ps}$. It can be shown that,

$$\mu_s = \overline{P}_{ss,sd} \frac{\lambda_{es}}{1 - \lambda_{ep}(\lambda_p / \mu_p)} I \left\{ 1 - \bar{a} \frac{\lambda_{ps}}{\mu_{ps}} \right\}. \quad (11)$$

Substituting with $\mu_{ps}$ from (9) in (11), it is noted that the resulting $\mu_s$ is independent of $\bar{a}$,

$$\mu_s = \overline{P}_{ss,sd} \frac{\lambda_{es}}{1 - \lambda_{ep}(\lambda_p / \mu_p)} I \left\{ 1 - \frac{P_{ps,pd}\overline{P}_{ps,ss} \lambda_{ep} \frac{\lambda_p}{\mu_p}}{\overline{P}_{ss,pd} \frac{\lambda_{es}}{1 - \lambda_{ep}(\lambda_p / \mu_p)} I} \right\}. \quad (12)$$

This can be explained as follows. When $a$ increases, the first term in the brackets of (7), $a\Pr(Q_{ps} \neq 0)$, would increase, and the probability of $Q_{ps}$ being empty, would decrease, which is the second term in the brackets of equation (7). So, changing the access probability $a$ has no effect on the SU service rate, $\mu_s$, in dominant system (*I*). From (1), (10), (12), the stability region of system (*I*) would be bounded by:

$$R_I = \left\{ \begin{array}{c} (\lambda_p, \lambda_s): \lambda_p < \mu_p, \\ \lambda_p < \frac{\bar{a}\,\overline{P}_{ss,pd}}{P_{ps,pd}\overline{P}_{ps,ss} \lambda_{ep}} \frac{\lambda_{es}}{1 - \lambda_{ep}(\lambda_p / \mu_p)} I \mu_p, \\ \lambda_s < \overline{P}_{ss,sd} \frac{\lambda_{es}}{1 - \lambda_{ep}(\lambda_p / \mu_p)} I \left\{ 1 - \frac{P_{ps,pd}\overline{P}_{ps,ss} \lambda_{ep} \frac{\lambda_p}{\mu_p}}{\overline{P}_{ss,pd} \frac{\lambda_{es}}{1 - \lambda_{ep}(\lambda_p / \mu_p)} I} \right\} \end{array} \right\}. \quad (13)$$

In Fig. 2, the achievable stable throughput region of dominant system (*I*) is plotted. Hereafter, the system parameters are chosen as follows: $\overline{P}_{ps,pd} = 0.3$, $\overline{P}_{ps,ss} = 0.4$, $\overline{P}_{ss,pd} = 0.7$, and $\overline{P}_{ss,sd} = 0.7$. The arrival rates of the two battery queues are as follow: $\lambda_{ep} = 0.6$, $\lambda_{es} = 0.6$. By examining the horizontal axis ($\lambda_p$) it is clear that the maximum sustainable $\lambda_p$ decreases as $a$ increases, where the SU tends to serve its data queue at the expense of the relay queue. First, as $\lambda_p$ increases, probability of PU being idle, $I$, decreases and $\Pr(Q_{es} \neq 0)$ increases with the same amount as $I$. From (11), the maximum sustainable $\lambda_s$ decreases as $\lambda_{ps}$ increases. Second, as $\lambda_p$ increases, $\Pr(Q_{es} \neq 0)$ reaches its maximum of unity, probability of PU being idle, $I$, decreases and $\lambda_{ps}$ increases. From (11), the rate of decrease of $\lambda_s$ would be faster as $\lambda_{ps}$ increases and $I$ decreases. This occurs when $\mu_{es} = 1 - \lambda_{ep}(\lambda_p/\mu_p) = \lambda_{es}$ and $\lambda_p = 0.25$. Third, the maximum sustainable $\lambda_p$ from (3) is $\mu_p = 0.34$.

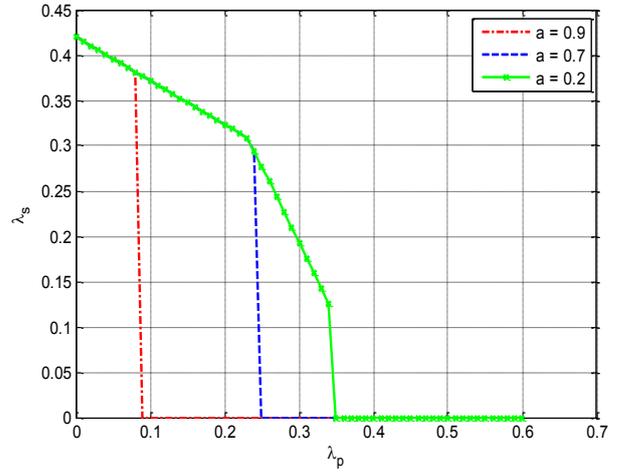

Fig. 2. Stable throughput region for dominant system (*I*) for different values of service probability $a$ ($\lambda_{ep} = 0.6, \lambda_{es} = 0.6$)

### B. Dominant System (II)

In this dominant system, dummy packets transmission from $Q_{ps}$ makes the service rate of the SU data queue, $Q_s$, decoupled from the state of $Q_{ps}$. A packet is served from $Q_s$ with probability $a$ if the PU is idle, channel between SU source and destination not in outage, and $Q_{es}$ not empty. For the SU data queue service rate,

$$\mu_s = \overline{P}_{ss,sd} \frac{\lambda_{es}}{1 - \lambda_{ep}(\lambda_p / \mu_p)} I\, a. \quad (14)$$

From (6) and with the relations $\Pr(Q_s \neq 0) = \lambda_s/\mu_s$ and $\Pr(Q_s = 0) = 1 - \lambda_s/\mu_s$, it can be verified that

$$\mu_{ps} = \overline{P}_{ss,pd} \frac{\lambda_{es}}{1 - \lambda_{ep}(\lambda_p / \mu_p)} I \left\{ 1 - a \frac{\lambda_s}{\mu_s} \right\}. \quad (15)$$

Substituting (14) in (15),

$$\mu_{ps} = \overline{P}_{ss,pd} \frac{\lambda_{es}}{1 - \lambda_{ep}(\lambda_p / \mu_p)} I \left\{ 1 - \frac{\lambda_s}{\overline{P}_{ss,sd} \frac{\lambda_{es}}{1 - \lambda_{ep}(\lambda_p / \mu_p)} I} \right\}. \quad (16)$$

It is noted that as $\bar{a}$ increases, the first term in the brackets of (6), $\bar{a}\Pr(Q_s \neq 0)$, would increase, and the probability of $Q_s$ being empty would decrease, which is the second term in the brackets of (6). So, changing the access probability $\bar{a}$, has no effect on the SU relay queue service rate, $\mu_{ps}$, in the dominant system (*II*). For the SU relay queue stability, with $\mu_{ps}$ derived in (16), it can be verified that,

$$\lambda_{ps} < \mu_{ps},$$

$$\lambda_p < \frac{\mu_p / \lambda_{ep}}{\overline{P}_{ss,pd} \Pr(Q_{es} \neq 0) + P_{ps,pd}\overline{P}_{ps,ss}} \times \left( \overline{P}_{ss,pd} \Pr(Q_{es} \neq 0) - \frac{\lambda_s}{C} \right), \quad (17)$$

where $C = \frac{\bar{P}_{ss,sd}}{\bar{P}_{ss,pd}}$. From (1), (14), (17), the stability region of dominant system (*II*) would be bounded by

$$R_2 = \begin{Bmatrix} (\lambda_p, \lambda_s): \lambda_p < \mu_p, \\ \lambda_s < \bar{P}_{ss,sd} \frac{\lambda_{es}}{1 - \lambda_{ep}(\lambda_p/\mu_p)} I a, \\ \lambda_p < \frac{\mu_p/\lambda_{ep}}{\bar{P}_{ss,pd} \Pr(Q_{es} \neq 0) + P_{ps,pd}\bar{P}_{ps,ss}} \\ \times \left( \bar{P}_{ss,pd} \Pr(Q_{es} \neq 0) - \frac{\lambda_s}{C} \right) \end{Bmatrix}. \quad (18)$$

Fig. 3 depicts the stability region under dominant system (*II*) for different values of service probability $a$. As shown in (16), the service rate of the relay queue, $\mu_{ps}$ is independent of $a$. The PU maximum allowable arrival rate ($\lambda_p$) depends only on $Q_p$ service rate, $\mu_p$, from (3) and $Q_{ps}$ service rate, $\mu_{ps}$, from (16). So, the PU maximum allowable arrival rate ($\lambda_p$) rate is independent of $a$. By observing the vertical axis ($\lambda_s$), the maximum sustainable ($\lambda_s$) increases as $a$ increases, where the SU tends to serve $Q_s$ with higher service probabilities.

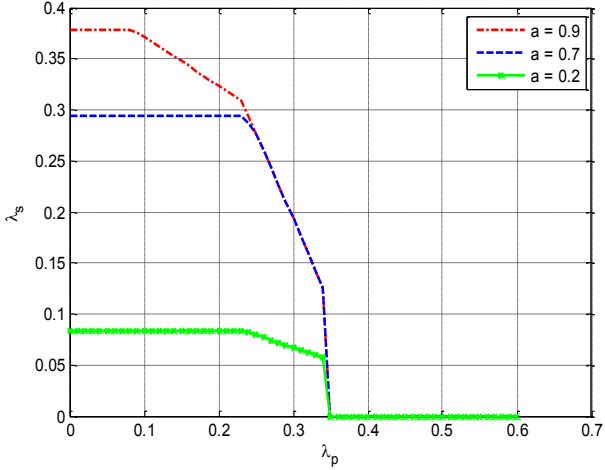

Fig. 3. Stable throughput region for dominant system (*II*) for different values of service probability $a$ ($\lambda_{ep} = 0.6, \lambda_{es} = 0.6$)

## IV. NUMERICAL RESULTS

In this section, the performance of the cooperative cognitive system under the energy constraint is investigated. The stability region of the discussed system is defined as the union of the two stable throughput regions of dominant system (I) and dominant system (II), $R = R_1 \cup R_2$. Hereafter, the system parameters are chosen as follows: $\bar{P}_{ps,pd} = 0.3$, $\bar{P}_{ps,ss} = 0.4$, $\bar{P}_{ss,pd} = 0.7$, and $\bar{P}_{ss,sd} = 0.7$. In Fig. 4, the stability region of the overall system, R, is plotted. It is worth mentioning that, in R the maximum allowable rates for the PU and SU are independent of $a$ since R is the union of the two dominant systems, $R_1 \cup R_2$, over all values of $a$. In Fig. 5, the system without energy constraint is compared to energy constrained system with $\lambda_{es} = 0.5$ and $\lambda_{ep} = 1$, i.e. energy constraint on the SU only. With the arrival rate at the SU battery queue is halved, the maximum sustainable SU throughput is ($\lambda_{es}\bar{P}_{ss,sd}$) = 0.35. It is intuitive that the energy constrained system ($\lambda_{es} = 0.5$) is always lower than the original system ($\lambda_{es} = 1$). This is true up to certain $\lambda_p$, after which the boundaries of the two stability regions coincide. This can be understood as from (8), the probability of $Q_{es}$ being non-empty is ($\lambda_{es}/I$). With an increasing $\lambda_p$, the probability that PU is idle decreases, and the probability of $Q_{es}$ being non-empty gradually approaches one. This makes the energy limited system act as the original system since in both systems $\Pr(Q_{es} \neq 0) = 1$. This value of $\lambda_p$ is reached when $\lambda_{es} = I = 1 - (1)(\lambda_p/\mu_p) = 0.5$ and $\mu_p = 0.58$ from (3), so, $\lambda_p = 0.29$.

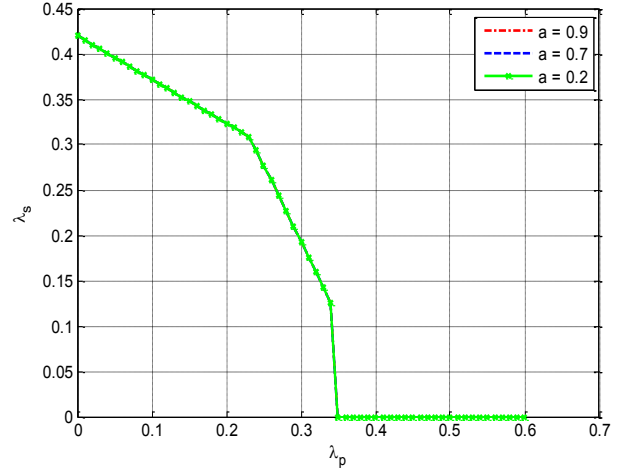

Fig. 4. Stable throughput region of the overall system ($\lambda_{ep} = 0.6, \lambda_{es} = 0.6$)

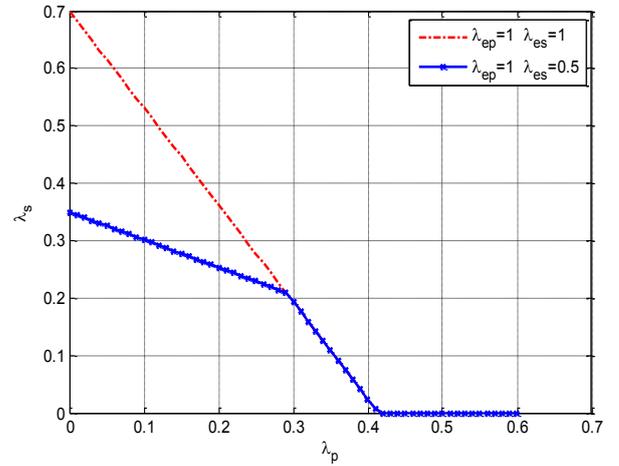

Fig. 5. Stable throughput region of the overall system with SU energy harvesing

In contrast to Fig. 5, system with energy constraint on the PU only is showed in Fig. 6. The energy constraint at PU is as, $\lambda_{ep} = 0.6$ and no constraint on SU since $\lambda_{es} = 1$. The influence of $\lambda_{ep}$ is directly obtained from (3), where, PU service rate, $\mu_p$, decreases as $\lambda_{ep}$ decreases. For the stability of $Q_p$, the maximum allowable PU arrival rate is also decreased to maintain the stability condition represented in (1). This cut in the horizontal axis ($\lambda_p$) occurs when the PU arrival rate equals the PU service rate, $\lambda_p = \mu_p = 0.6 \times 0.58 = 0.34$.

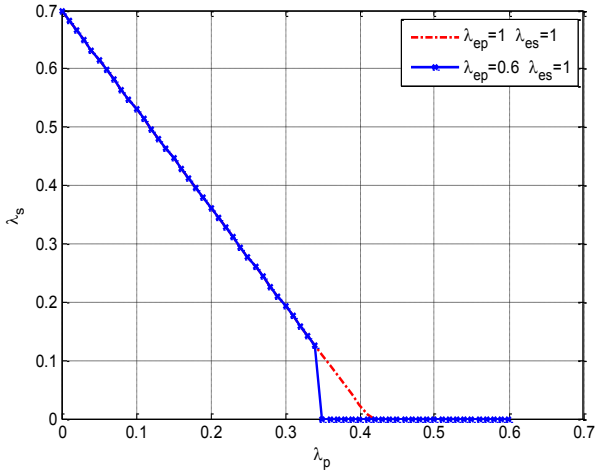

Fig. 6. Stable throughput region of the overall system with PU energy harvesing

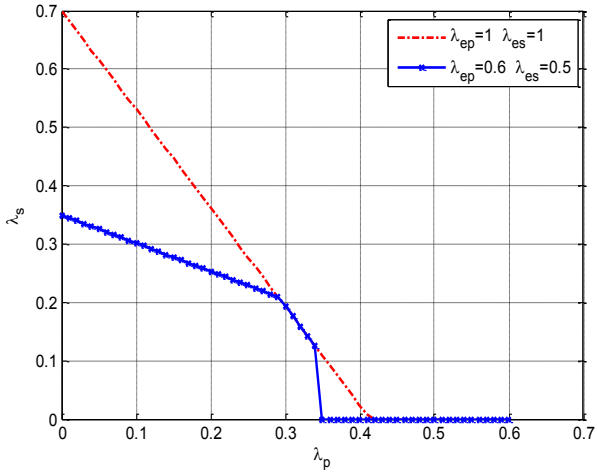

Fig. 7. Stable throughput region of the overall system with PU and SU energy harvesing

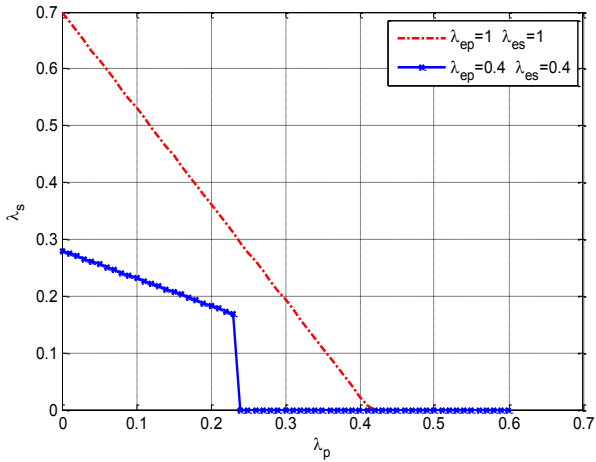

Fig. 8. Stable throughput region of the overall system with PU and SU energy harvesing

Combing the effect of energy limitation on PU and SU, shrinks the stability region in both directions (for PU and SU) as in Fig. 7, it is seen that the two boundaries coincide when the PU activity increases and SU battery queue is able to cover transmission on the available time slots. In Fig. 8, this mentioned coincidence does not exist since, with more severe energy limitation, the PU probability of being idle is higher and no sufficient energy at the SU to exploit these idle time slots.

*Noncooperative energy harvesting system*

For the sake of comparison, we introduce the energy harvesting cognitive radio system without cooperation between PU and SU. For $Q_p$, a packet is serviced if it is successfully decoded by the PU destination only. With the same derivations of service rates of battery queues and $I$, we can get

$$\mu_p = \overline{P}_{ps,pd}\, \lambda_{ep}, \qquad (19)$$

$$\mu_s = \overline{P}_{ss,sd}\, \frac{\lambda_{es}}{1 - \lambda_{ep}\,(\lambda_p / \mu_p)}\, I. \qquad (20)$$

The stability regions of the two systems (with and without cooperation) are plotted in Fig. 9 and Fig. 10 with $\lambda_{ep} = 0.5$, $\lambda_{es} = 0.8$ and $\lambda_{ep} = 0.5$, $\lambda_{es} = 0.6$, respectively.. It is clear that for the two systems, the maximum sustainable SU throughput is $(\lambda_{es}\overline{P}_{ss,sd})$. Cooperative system provides better SU throughput for all possible values of $\lambda_p$ also, the maximum sustainable PU arrival rate, $\lambda_p$, is larger in the cooperative system. This means that, cooperation between PU and SU is beneficial for both users.

Fig. 9 and Fig. 10 show that non-cooperative energy harvesting system is better than cooperative energy harvesting system for low values of $\lambda_p$ in terms of SU throughput. At certain $\lambda_p$, the two systems intersect providing the same SU allowable throughput, after which, the cooperative system is better in terms of SU throughput and maximum allowable PU arrival rate.

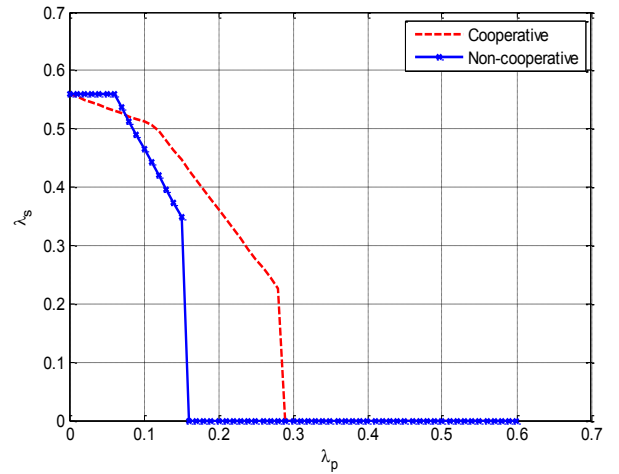

Fig. 9. Stable throughput region of the overall cooperative system and noncooperative system ($\lambda_{ep} = 0.5, \lambda_{es} = 0.8$)

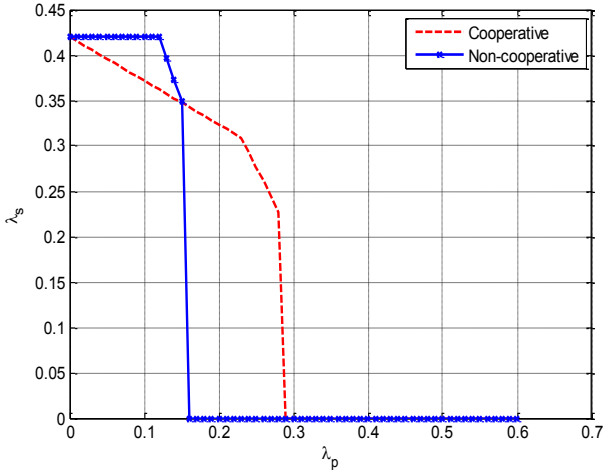

Fig. 10. Stable throughput region of the overall cooperative system and noncooperative system ($\lambda_{ep} = 0.5$, $\lambda_{es} = 0.6$)

To obtain the value of $\lambda_p$ after which the energy harvesting cooperative system becomes superior to the non-cooperative system, we would equate $\mu_s$ from (12) with $\mu_s$ from (20). After some algebraic manipulation, we get

$$\Lambda_p = \frac{1 - \lambda_{es}}{D}, \qquad (21)$$

where $\Lambda_p$ is the PU arrival rate at which the two systems (cooperative and non-cooperative) are with the same SU allowable throughput, and D is a constant function of the channel outages probabilities

$$D = \frac{1}{\bar{P}_{ps,pd}} - \frac{P_{ps,pd}\bar{P}_{ps,ss}}{\bar{P}_{ss,pd}(\bar{P}_{ps,pd} + P_{ps,pd}\bar{P}_{ps,ss})} . \qquad (22)$$

From (22), it is clear that the value, $\Lambda_p$, at which the two systems intersect is a function only of the system channel parameters and $\lambda_{es}$. With the mentioned values of $\lambda_{es}$, $\Lambda_p = 0.075$ in Fig. 9 and $\Lambda_p = 0.15$ for Fig. 10. It is clear that this value of $\Lambda_p$ decreases as $\lambda_{es}$ increases, and also, it is independent of $\lambda_{ep}$. This result can be interpreted as follows; the SU with higher energy levels at its battery is supposed to has better opportunity to cooperate with the PU and leverage the better channel conditions between the SU and PU destination.

## V. CONCLUSION

This paper studied the stability region of energy harvesting cooperative cognitive radio network, where the nodes have rechargeable energy sources. The rechargeable energy sources are modeled as queues; along with the data queues at each node they form an interacting system of queues. The stability region of this system is characterized using the dominant system approach. A comparison between the cooperative and non-cooperative systems is introduced for different energy harvesting rates. The stability regions of the two systems showed that, under certain channel outage probabilities, the non-cooperative system outperforms the cooperative system for lower PU arrival rate. After certain PU arrival rate, the situation is reversed and the cooperative system is dominant. The PU arrival rate, at which the two regions intersect, is deduced in terms of channel outage probabilities.